\documentclass[]{spie}  
\usepackage[]{graphicx}

\title{Quality functions in community detection} 

\author{Santo Fortunato
\skiplinehalf
Complex Networks Lagrange Laboratory (CNLL), ISI Foundation, Torino, Italy}

\authorinfo{E-mail: fortunato@isi.it, telephone: +39-011-6603555.}

\begin{document} 
  \maketitle 

\begin{abstract}
Community structure represents the local organization 
of complex networks and the single most important feature to 
extract functional relationships between nodes. In the last years, the 
problem of community detection has been reformulated in terms of the 
optimization of a function, the Newman-Girvan modularity, that is supposed to 
express the quality of the partitions of a network into communities.
Starting from a recent critical survey on modularity optimization, pointing out 
the existence of a resolution limit that poses severe limits to its applicability,
we discuss the general issue of the use of quality functions in
community detection. Our main conclusion is that quality functions 
are useful to compare partitions with the same number of modules,  
whereas the comparison of partitions with different numbers of modules is not straightforward and
may lead to ambiguities.
\end{abstract}



\section{INTRODUCTION}
\label{sect:intro}  

The importance of networks in modern science 
can hardly be underestimated.
The network representation, where the elementary units of a system become 
vertices connected by relational links, has proved very 
successful to understand the structure and dynamics of social,
biological and technological systems, with the big advantage 
of a simple level of description~\cite{bara02,mendes03,Newman:2003,psvbook,vitorep}.

The structure of a network can be studied at the global level, focusing on
statistical distributions of topological quantities, like degree, clustering coefficient, 
degree-degree correlations, etc., or at the local level, disclosing how nodes are organized
according to their specific features.
Topologically, such local organization of the nodes is revealed by the existence 
of subsets of the network, called communities or modules, with many links between 
nodes of the same subset and only a few between nodes of different subsets. 
Communities can be considered as relatively independent units of the whole 
network, and identify classes of nodes with common features and/or 
special functional relationships. For instance, 
communities represent sets of pages dealing with the same topic in the World Wide Web~\cite{Flake:2002}, 
groups of affine individuals in social networks~\cite{Girvan:2002,Lusseau:2003, Adamic:2005}, 
compartments in food webs~\cite{foodw1,foodw2}, etc.

The problem of identifying communities in networks has recently turned into an optimization 
problem, involving a quality function introduced by Newman and Girvan~\cite{Newman:2004b}, called 
modularity. This function should evaluate the ``goodness'' of a partition of a network
into communities. The general idea is that a subset of a network is a module 
if the number of internal links exceeds the
number of links that one expects to find in the subset if the network were random. 
If this is the case, one infers that
the interactions between the nodes of the subset are not random, which means that 
the nodes form an organized subset, or module.
Technically, one compares 
the  number of links inside a given module
with the expected number of links in a randomized version of the network that keeps the same
degree sequence. The partition is the better, the larger the excess of links in each module
with respect to the random case.
In this way, the best partition of the network is the one that maximizes
modularity. Optimizing modularity is a challenging task, as the number of possible 
partitions of a network increases at least exponentially with its size. Indeed, it has been recently 
proven that modularity optimization is an NP-complete problem~\cite{brandes}, so one has to give up
the ambitious goal of finding the true optimum of the measure and content oneself with 
methods that deliver only approximations of the optimum, like greedy agglomeration~\cite{Newman:2004c,Clauset:2004},
simulated annealing~\cite{Guimera:2004,Guimera:2005,Reichardt:2006}, extremal optimization~\cite{Duch:2005} 
and spectral division~\cite{Newman:2006}. 

We believe that the scientific community has been a little too fast in adopting modularity 
optimization as the most promising method to detect communities in networks. Indeed, 
all research efforts focused on the creation of an effective algorithm to 
find the modularity maximum, without preliminary investigations on the measure itself
and its possible limitations. Only recently a critical examination has been performed, 
revealing that modularity has an intrinsic resolution scale, depending on the size of the system,
so that modules smaller than that scale may not be resolved~\cite{fortunato:2007}. 
This represents a serious problem for the 
applicability of modularity optimization, especially when the network at study is large. The existence of this bias
has also been revealed~\cite{kumpula:2007} in the Hamiltonian formulation of modularity 
introduced by Reichardt and Bornholdt~\cite{Reichardt:2006}, that leaves some
freedom in the criterion determining whether a subset is a module or not.
Other doubts about modularity and its applicability were raised before the discovery 
of the resolution limit~\cite{Muff:2005}.

In our opinion, the problems of modularity optimization call for a  
debate about the opportunity to use quality functions 
to detect communities in networks. This general issue,
which has never been discussed in the literature on community detection, is the subject 
of this paper. We start with an analysis of modularity, where we illustrate its
features as well as its limits. Such analysis is a valuable guide to 
uncover the possible problems that arbitrary quality functions may have, to understand what    
determines these problems and what can be done to solve them. We will see that, while   
it is easy to define a quality function within classes of partitions 
with the same number of modules, it is not clear how to compare network splits that differ
in the number of modules. 

This paper reproposes some results of the recent work~\cite{fortunato:2007}, carried out in collaboration 
with Dr. Marc Barth\'elemy, integrating them with new material and discussion.
In Section 2 we introduce and analyze the modularity
of Newman and Girvan; in Section 3 we deal with the general issue of quality functions and
their applicability; our conclusions are summarized in Section 4.

\section{Modularity optimization and its problems}

\subsection{Definition and properties}

The modularity of a partition in modules of a network with N nodes and L links 
can be written in different equivalent ways.
We stick to the following expression 
\begin{equation}
Q=\sum_{s=1}^{m}\Big[\frac{l_s}{L}-\left(\frac{d_s}{2L}\right)^2\Big],
\label{eq:mod}
\end{equation}
where the sum is over the $m$ modules of the partition, $l_s$ is the
number of links inside module $s$ and $d_s$ is the total degree of the nodes in module
$s$. 

Any method of community detection is bound to start from stating what a community is.
In the case of modularity the definition of community is revealed by
each summand of Eq.~(\ref{eq:mod}), where we distinguish two terms, $A_s=l_s/L$ and $B_s=(d_s/2L)^2$. 
The term $A_s$ is the fraction of links connecting pairs of nodes belonging to module $s$, whereas
$B_s$ represents the fraction of links that one would expect to find inside that module if links were
placed at random in the network, under the only constraint that the
degree sequence coincides with that in the original graph. In this way, if $A_s$ exceeds $B_s$,
the subset $s$ of the network is indeed a module, as it presents more links than expected
by random chance. The larger this excess of links, the better defined the module. We conclude that,
within the modularity framework, a subgraph ${\cal S}$ with $l_s$ internal links and total degree $d_s$
is a module if
\begin{equation}
\frac{l_s}{L}-\left(\frac{d_s}{2L}\right)^2>0.
\label{eq2}
\end{equation}
Starting from this definition, Newman and Girvan deduced 
that the overall quality of the partition
is given by the sum of the qualities of the individual modules, which is not straightforward, 
as we shall see in the next section. 

If all subsets $s$ of the partition are modules, in the sense specified by Eq.~(\ref{eq2}),
the modularity of the partition is positive, i.e. $Q>0$. On the other hand, modularity is a bounded 
function. Since each summand cannot be larger than the term $A_s$, one has
\begin{equation}
Q=\sum_{s=1}^{m}\Big[\frac{l_s}{L}-\left(\frac{d_s}{2L}\right)^2\Big]
\leq\sum_{s=1}^{m}\frac{l_s}{L}=\frac{1}{L}\sum_{s=1}^{m}l_s\leq 1.
\label{eq3}
\end{equation}
In this way, for any network, $Q$ has a well defined maximum.
Since the partition into a single module, i.e. the network itself,
yields $Q=0$, as in this case $l_1=L$ and $d_1=2L$, we conclude that 
the maximal modularity is non-negative. Practical applications
suggest that partitions with modularity values of about $0.3-0.4$  
correspond to well defined community structures. However, these numbers should be 
taken with a grain of salt: the modularity maximum usually increases with
the size of the network at study, so it is not meaningful to compare the quality of 
partitions of networks of very different size based on the relative values of $Q$. Moreover,
the modularity maximum of a network yields a meaningful partition only if 
it is appreciably larger than the modularity maximum expected for a random graph of the same
size~\cite{Bornholdt:2006}, as the latter may attain very high values,
due to fluctuations~\cite{Guimera:2004}.

\subsection{The resolution limit}

Let us consider two subsets ${\cal S}_1$ and ${\cal S}_2$ of a network. The total degree 
of each subset is $d_1$ for ${\cal S}_1$ and $d_2$ for ${\cal S}_2$. We want to calculate
the expected number of links connecting the two subsets in the null model 
chosen by modularity, i.e. a random network with the same size and degree sequence of the 
original network. 

By construction, the total degrees of ${\cal S}_1$ and ${\cal S}_2$
in the randomized version of the network will be the same\footnote{For the sake of precision, one should say that
their expectation values over the class of possible randomizations
of the network are the same, but this does not affect our argument.}. Each link of the network
consists of two halves, or stubs, originating each from either of the nodes connected by the link.
The probability that one of the stubs originates from a node of ${\cal S}_1$ is $p_1=d_1/2L$; similarly,
the probability that the other stub of the link originates from a node of ${\cal S}_2$ is $p_2=d_2/2L$.
So, the probability that a link of the random network joins a node of ${\cal S}_1$ with a node of 
${\cal S}_2$ is $2p_1p_2=d_1d_2/2L^2$, where the factor two is due to the symmetry of the link with respect to the 
exchange of its two extremes. As there are $L$ links in total, the expected number $l_{12}$ of links 
connecting ${\cal S}_1$ and ${\cal S}_2$ is
\begin{equation}
l_{12}=2p_1p_2\,L=\frac{d_1d_2}{2L}.
\label{eq4}
\end{equation}
Now we notice something interesting. In all our discussion we set no constraint on the
parameters $d_1$, $d_2$ and $L$ other than the trivial conditions $d_1\leq 2L$ and $d_2\leq 2L$,
as the total degree of either subset cannot exceed the total degree of the network by construction.
In particular, $d_1$ and $d_2$ could be much smaller than $L$, and $l_{12}$ could be smaller than one.
To simplify the discussion, we assume that both ${\cal S}_1$ and ${\cal S}_2$ have equal
total degree, i.e. $d_1=d_2=d$. In this case, the condition $l_{12}<1$ implies 
$d^2/2L<1$, or, equivalently,
\begin{equation}
d<d_{lim}=\sqrt{2L}.
\label{eq5}
\end{equation}
In this way, if the total degree of either subset is smaller than $d_{lim}$, the expected number of 
interconnecting links in the random network would be less than one, so if there is even a single
link between them in the original network, modularity would merge ${\cal S}_1$ and ${\cal S}_2$ in the same module.
This is because the two subsets would appear more connected than expected by random chance.
We have made no hypothesis on the subsets: the number of nodes in them 
does not play a role in our argument, as it does not enter the definition of modularity, 
as well as the distribution of links inside them. For all we know,
${\cal S}_1$ and ${\cal S}_2$ could even be two complete graphs, or cliques, which represent the 
most tightly connected subsets one can possibly have, as every node of a clique 
is connected with all other nodes. We found that, regardless of that, optimizing 
modularity would make them parts of the same module, even if they appear very weakly 
connected, since they share only one link.

\begin{figure}
\begin{center}
\begin{tabular}{c}
\includegraphics[height=10cm]{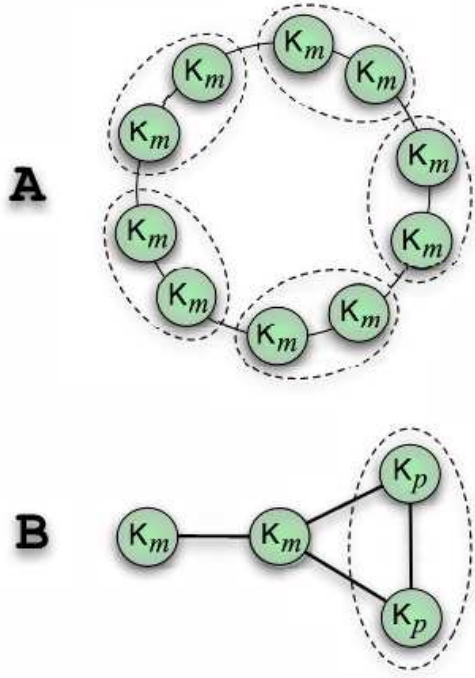}
\end{tabular}
\end{center}
\caption{\label{fig1} Resolution limit of modularity optimization. A. The illustrated network  
is a set of complete graphs with $m$ nodes each, arranged on a ring-like structure, such that
two consecutive cliques are connected by a single link. If the number of cliques is larger than $\sqrt{L}$,
modularity optimization will not be able to resolve the individual cliques, but it would merge them in 
groups of two or more (indicated by the dotted line). B. The network in the figure is made out of two 
pairs of identical cliques, with $m$ and $p<m$ nodes, respectively. If $m$ is large enough with respect to $p$,
the best partition of the network according to modularity is a split in three modules, with the two smaller 
cliques combined in a single module (indicated by the dotted line).}
\end{figure}

Some striking consequences of this finding are shown in Fig.~\ref{fig1}, where we  
present two schematic examples. In the first example, we consider a network made out
of $K_m$ cliques, i.e. graphs with $m$ nodes and  $m(m-1)/2$ links. Each clique is connected
to two others by one link, forming a ring-like structure (Fig.~\ref{fig1}A). We have $n$ cliques,
so that the network has a total of $N=nm$ nodes
and $L=nm(m-1)/2+n$ links. The natural community structure of the network is represented by 
the partition where each module corresponds to a single clique. The modularity 
$Q_{single}$ of this partition equals
\begin{equation}
Q_{single}=1-\frac{2}{m(m-1)+2}-\frac{1}{n},
\label{eq6}
\end{equation}
and we would expect that 
$Q_{single}$ is the maximum modularity for this network. If this is true, 
$Q_{single}$ should be larger than the $Q$-value of any other partition of the network.
Let us consider the partition where the modules are
pairs of consecutive cliques, delimited by the dotted lines in Fig.~\ref{fig1}A. The modularity
$Q_{pairs}$ of this partition is
\begin{equation}
Q_{pairs}=1-\frac{1}{m(m-1)+2}-\frac{2}{n}.
\label{eq7}
\end{equation}
The condition $Q_{single}>Q_{pairs}$ is equivalent to 
\begin{equation}
m(m-1)+2>n,
\label{eq8}
\end{equation}
which is not always true, as the variables $m$ and $n$ are independent of each other,
and therefore it is possible to choose their values such that the inequality~(\ref{eq8})
is not satisfied. For instance, for $m=5$ and $n=30$, $Q_{single}=0.876$ and
$Q_{pairs}=0.888>Q_{single}$. So, in this case, the modularity maximum would not 
correspond to the natural community structure of the network. Likewise, in the
example illustrated in Fig.~\ref{fig1}B, the network includes four cliques: two with $m$
nodes each, the other two with $p<m$ nodes. By choosing $m=20$ and $p=5$, the modularity 
maximum is attained for the partition in three modules illustrated in the figure,
and the two smaller cliques are not resolved.

The examples we considered are very different from real networks,
but the conclusion is absolutely general: modularity increases by merging subsets of nodes with total degree
of the order of $d_{lim}$ or smaller. Consequently, even evident community structures may not be resolved,
if the size (in degree) of the modules lies below the resolution limit $d_{lim}$. This is actually only 
part of the story: pairs of communities may be merged even if they differ considerably in size, 
as long as the condition in Eq.~(\ref{eq4}) is satisfied. In the latter scenario, one of the two subsets can
in principle be much bigger (in degree) than $d_{lim}$. 

From our discussion it is clear that the resolution limit of modularity is induced by the 
null model adopted in this framework, i.e. by the fact that the network at study is compared 
with a randomized version of it. In the random network, each node has the same probability to 
be attached to any other node, as long as its degree is kept constant, which means that one makes the implicit assumption that 
each node has a complete information about the network. This is certainly not true in general, 
especially for large networks. Instead, every node usually interacts with a limited number of peers,
ignoring the rest. Community structure depends on the local organization of groups of nodes, 
it has nothing to do with the network at large. The possibility to introduce a local concept
of modularity has been explored~\cite{Muff:2005}.

\subsection{Evolving networks}

From Eq.~(\ref{eq5}), we see that $d_{lim}$ increases with the total number $L$ of links
of the network. Therefore, the larger the network, the more likely it is 
to have to do with community structures that modularity optimization is not able to resolve. 
But this also has another consequence, that we discuss in this section.

Most real networks, if not all, are dynamic structures, that change  
considerably in time. For example, the graph of the World Wide Web, where nodes and links identify 
URLs and hyperlinks, respectively, has undergone an exponential growth in the fifteen years 
elapsed since its birth. 
The analysis of community structure
is usually performed on static snapshots of evolving graphs, mostly due to the lack of data.
But in principle we could think of detecting the communities of the network along its time evolution.
This is very instructive, as one could monitor how nodes organize among themselves in time.

\begin{figure}
\begin{center}
\begin{tabular}{c}
\includegraphics[height=10cm]{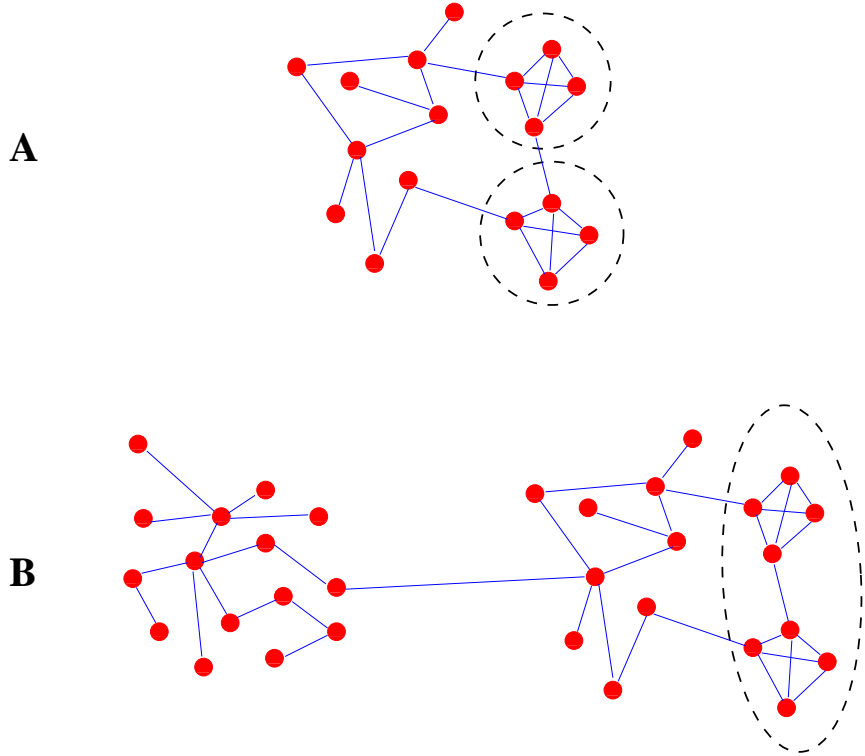}
\end{tabular}
\end{center}
\caption{\label{fig30} Modularity optimization in evolving networks. A. The two cliques on the right 
are correctly identified as separate communities. B. At a later time, the network becomes much larger due to the
merge with another network. The increased size of the resulting network increases the resolution limit and modularity
optimization may now merge the two cliques, even if, from the point of view of topology, nothing changed in the structure
of the cliques nor in their neighborhood.}
\end{figure}

In this context, the dependence of $d_{lim}$ on $L$ can lead to strange results, as illustrated in Fig.~\ref{fig30}. 
We have a network with a pair of weakly connected cliques,  
the connection being represented by a single link. Let us suppose that 
the modularity maximum of the network corresponds to a partition where 
the two cliques are separated (Fig.~\ref{fig30}A). At some point, the network 
merges with another network (we could think of different friendship circles
with two people belonging to different circles 
that get in touch and become friends, joining the two communities).  
Now the system has a larger size and the resolution limit rises accordingly, so that
the two cliques may not be considered as separate entities and could be merged in a single module (Fig.~\ref{fig30}B).
This is odd, as the fusion of the two networks does not seem to affect the
mutual relationships of the nodes belonging to the cliques, nor their interactions with 
neighboring nodes, so we would say that the local organization of that
part of the network was not affected by its evolution. The conclusion is that the answer obtained from
modularity optimization may change in time, even when the local organization of the network is preserved. 
This is because the scale at which we are exploring the system changes in the course of its evolution, independently of the 
network structure. 

\section{Searching for a quality function}

\subsection{The quality of a partition}

The first important issue to address is the definition of community.
Let ${\cal G}$ be a subgraph of a network. A general
condition for ${\cal G}$ to be a community can be expressed as
\begin{equation}
Q_{\cal G}(l^{in}_{\cal G},l^{out}_{\cal G},n_{\cal G},...,N,L)>0,
\label{eq9}
\end{equation}
where $Q_{\cal G}$ is a function of some topological properties,
like the number $l^{in}_{\cal G}$ ($l^{out}_{\cal G}$) of internal (external) links of $\cal G$, the number $n_{\cal G}$
of nodes of ${\cal G}$, the total number $N$ of nodes and $L$ of links of the whole network, etc.
For instance, in the case of Newman-Girvan modularity, the above condition has the expression of 
Eq.~(\ref{eq2}). We can reasonably assume that, the larger
the value of $Q_{\cal G}$ (as long as it is positive), the more 
``community-like'' is $\cal G$. This is certainly the best thing to do when one 
wishes to qualify a subgraph of a network as a community. 

But the problem of community detection is more 
complex. Ideally, we would like to find a partition in a number of ``good'' parts.
Any algorithm has to say how many communities there are in the network and assign each node
to its community. So, it is necessary to evaluate the goodness of network partitions, to be able to discriminate
between them. That is what quality functions are needed for. 
The crucial question is: 

{\it what is the best way to qualify the partition of a network, based on 
the function $Q_{\cal G}$ expressing the quality of a single subgraph?}

The solution proposed by Newman and Girvan for their modularity is to sum the qualities of all 
subgraphs of a partition (see Eq.~(\ref{eq:mod})). We remark that, while this looks like 
a reasonable option, it is neither the only possibility
nor necessarily the best one. For instance, one could consider the average value of the quality of a module 
in the partition, the product of the individual qualities, etc. From this point of view, there seems to be 
a substantial degree of arbitrariness in the definition of a quality function, that can be an arbitrary 
function ${\cal Q}(Q_{{\cal G}_1},Q_{{\cal G}_2}, ...,Q_{{\cal G}_m},m)$ of the individual qualities 
of the modules and their number $m$. 

We can try to limit this freedom by imposing some conditions on our quality function, so that it can best serve 
its main purpose, i.e. allowing for an objective comparison of network partitions. A first trivial condition
is that our ${\cal Q}$ should be an increasing function of the individual $Q_{\cal G}$, so that, the higher the quality
of the modules, the better the partition. Other constraints come when one considers the comparison of two
different network partitions. There are two possibilities:
\begin{itemize} 
\item{the numbers of modules of the partitions are equal;}
\item{the numbers of modules of the partitions are different.}
\end{itemize}

Let us suppose that we want to find the best partition of the network in $m$ modules. 
The problem is equivalent to having $N$ balls and $m$ boxes, where the balls represent the nodes of the network.
Each network split corresponds to a possible distribution of the balls inside the boxes. One can pass from every partition
to any other by moving balls to different boxes. Let us compare two partitions 
${\cal P}_1$ and ${\cal P}_2$ that differ from each other only
by shifting a single ball from box $i$ to box $j$. In this case, the partitions will only differ in the 
boxes $i$ and $j$, so we can neglect the others. From a topological point of view, the two configurations
are symmetric, and the only question is whether the node that makes the difference should stay 
in module $i$ or $j$.
Therefore, we need a function 
such that the difference between the 
qualities of ${\cal P}_1$ and ${\cal P}_2$ only involves the qualities of modules $i$ and $j$.
Possible {\it ansaetze} satisfying these conditions are the sum of $Q_{\cal G}$ over all modules, like in 
Newman-Girvan modularity, the product of $Q_{\cal G}$ over all modules, etc. 
We conclude that comparing the quality of partitions with a fixed number of modules is possible, and
that the modularity of Newman and Girvan, in spite of its problems,  may be good at that. 

\begin{figure}
\begin{center}
\begin{tabular}{c}
\includegraphics[height=7cm]{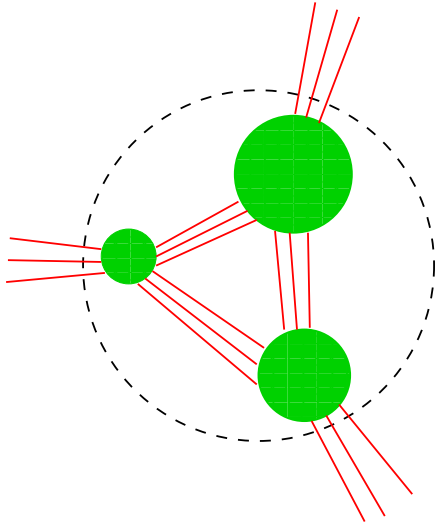}
\end{tabular}
\end{center}
\caption{\label{fig3} 
The puzzle of the number of modules. The circles are subsets of the network that satisfy the condition in Eq.~(\ref{eq9}), so they 
are modules. Their union, indicated by the dashed circular boundary, is itself a module. Which configuration is better?}
\end{figure}

We now examine the case in which the number of modules is different in the two partitions.
Here, it is no longer possible to transform a partition into another by simply shifting 
nodes and the main question is in how many classes it is appropriate to distribute the nodes.
This issue is illustrated schematically in Fig.~\ref{fig3}, where the full circles indicate 
three communities according to the definition of Eq.~(\ref{eq9}). Let us suppose that the
subset of the network represented by the nodes of the three modules and their links is 
as well a module according to Eq.~(\ref{eq9}). How can we say whether the nodes 
are organized in a single module or in the three smaller modules, based on the numbers expressing the qualities of each
module? Now the configurations are no longer symmetric, and we do not see a clear way to address the issue.
Moreover, the problem could be ill-defined, as it is possible that both configurations are meaningful, 
because they correspond to different hierarchies in the local organization of the network. 
The optimization of Newman-Girvan modularity would deliver either alternative, depending on the number of links 
of the subgraph as compared with the total number of links of the network.

\subsection{The ideal partition}

Solving the puzzle of the number of modules of the ``best'' partition of a network,
presented in the previous section, is equivalent to finding a prescription for a suitable quality 
funtion. To partially address this issue, 
we propose a criterion that a quality function should respect in order to be reliable.
The criterion involves the concept of ideal partition of a network, and is based on the fact that a good quality
function must attain its highest possible value in correspondence of this partition.
\begin{figure}
\begin{center}
\begin{tabular}{c}
\includegraphics[height=7cm]{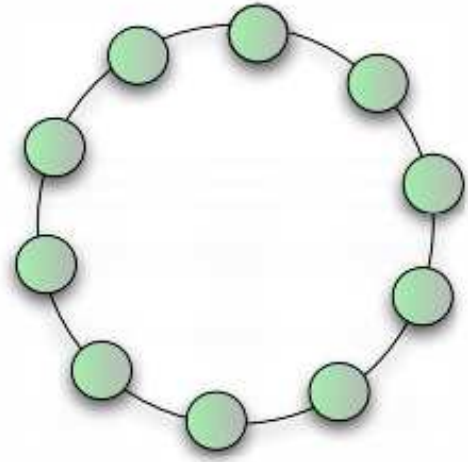}
\end{tabular}
\end{center}
\caption{\label{fig4}Ideal network partition. The circles represent cliques,  
each clique having a number of nodes equal to the closest integers to $d+1$, 
where $d$ is the average degree of the network.}
\end{figure} 

We start with a set of $N$ nodes and $L$ links. We want to distribute the links among the nodes in order to
build the ideal ``modular'' network. What kind of network is it? Intuitively, we expect that the network 
presents groups of nodes with the highest possible density of links between nodes of the same group, and the smallest 
possible number of connections between the groups. We assume 
that we have $m$ identical groups, for symmetry reasons.
The highest density of links inside each group is attained when the latter is a 
complete graph. The ideal configuration should have $m-1$ interconnecting links, which is the minimum number
of links necessary to keep the network connected. For the sake of symmetry, we instead use $m$ links, so that the
cliques can be arranged in the ring-like structure schematically illustrated in Fig.~\ref{fig4}.  
The average degree of the network is fixed by construction to the value $d=2L/N$, and all nodes
essentially have the same degree, with slight differences
depending on their being connected or not to nodes of a different group. In this way, the cliques comprise
$d+1$ nodes\footnote{The average degree $d$ is in general not integer. Therefore the statement
means that some cliques have $[d]+1$ and some others $[d]+2$ nodes, with $[d]$ the integer part of $d$,
so to respect the constraint on the total number of nodes and links of the network.}.
The number of cliques is then approximately $N/(d+1)$, which is an important constraint 
on the desired quality function, and a useful indication on how to group nodes into modules.

To test a quality function, one could identify the network partition that delivers the
highest possible value of the measure, and check whether it coincides with the ideal partition that we have derived, 
or in which respect it is different from it.

In the case of Newman-Girvan modularity, such partition can be easily determined~\cite{fortunato:2007}.
We proceed in two steps: first, we consider the maximal value $Q_M(m,L)$ of modularity for a partition into a fixed number $m$ of
modules; after that, we look for the number $m^\star$ that maximizes $Q_M(m,L)$. 
Again, the best configuration is the one with the smallest number of links connecting different modules. For simplicity we
shall assume that there are $m$ bridges between the modules, so that the network resembles the one 
in Fig.~\ref{fig4}. The modularity of such a network is
\begin{equation}
Q=\sum_{s=1}^{m}\Big[\frac{l_s}{L}-\left(\frac{2l_s+2}{2L}\right)^2\Big],
\label{eq10}
\end{equation}
where 
\begin{equation}
\sum_{s=1}^{m}l_s=L-m.
\label{eq11}
\end{equation}
The maximum is reached when all modules contain the same number of links, 
i.e. $l_s=l=L/m-1, \forall s=1,2,...,m$. Its value equals
\begin{equation}
Q_M(m,L)=m\Big[\frac{L/m-1}{L}-\left(\frac{L/m}{L}\right)^2\Big]=1-\frac{m}{L}-\frac{1}{m}.
\label{eq12}
\end{equation}
We have now to find the maximum of $Q_M(m,L)$ when the number of
modules $m$ is variable. For this purpose we treat $m$ as a real
variable and take the derivative of $Q_M(m,L)$ with respect to $m$
\begin{equation}
\frac{dQ_M(m,L)}{dm}=-\frac{1}{L}+\frac{1}{m^2}
\label{eq13}
\end{equation}
which vanishes when $m=m^\star=\sqrt{L}$. We conclude that the network with the highest possible modularity
comprises $\sqrt{L}$ modules, with each module consisting 
of about $N/\sqrt{L}$ nodes and $\sqrt{L}$ links. 
The resolution scale of modularity optimization
emerges at this stage, where we see that the best possible partition requires modules with total degree of about 
$d_{lim}=\sqrt{2L}$. The modules in general are not cliques, at variance with those of the 
ideal network partition. This is due to the fact that the number of nodes inside the modules 
does not affect the value of the modularity of a partition.

Modifications of modularity do not improve the situation. As an example,
we introduce a modified modularity $Q^A$, 
that differs from the original measure of Newman and Girvan 
in that the quality of the partition is not the sum, but the average value of the qualities of the modules. This is {\it a priori} 
a meaningful definition and its expression reads 
\begin{equation}
Q^A=\frac{1}{m}\sum_{s=1}^{m}\Big[\frac{l_s}{L}-\left(\frac{d_s}{2L}\right)^2\Big],
\label{eq14}
\end{equation}
where the symbols have the same meaning as in Eq.~(\ref{eq:mod}). Again, we wish to find the 
network partition that delivers the highest possible value for $Q^A$. The procedure adopted for
the original modularity applies in this case as well until Eq.~(\ref{eq12}), which now takes the form
\begin{equation}
Q^A_M(m,L)=\frac{1}{m}-\frac{1}{L}-\frac{1}{m^2},
\label{eq15}
\end{equation}
whose derivative with respect to $m$ is
\begin{equation}
\frac{dQ^A_M(m,L)}{dm}=-\frac{1}{m^2}+\frac{2}{m^3},
\label{eq16}
\end{equation}
which vanishes when $m=m^\star=2$. So, the ideal network partition for the new modularity $Q^A$ 
is a split in two communities of the same size, independently of the number of nodes and links of the network.
The resolution scale of $Q^A$ is then of the order of the size (in degree) of the two communities, which
is $L$. Because of that, the optimization of $Q^A$ delivers partitions in a small number of modules,
which means that the network is examined at a coarser level with respect to the original modularity and the situation is 
much worse than before.

\section{Conclusions}

Quality functions allow to convert the problem of community detection into an optimization
problem. This has big advantages, potentially, because one can exploit a wide variety of
techniques and methods developed for other optimization problems. In this paper we 
used the modularity of Newman and Girvan as a paradigm to discuss 
the problem of the definition of a quality function suitable 
for community detection. We have seen that modularity cannot scan the network below
some scale, and that this may leave small modules undetected, even when they are easily identifiable.
Moreover, the identification of modules may be affected by the time evolution of the network, due to 
the fact that modularity's resolution scale varies with the size of the network.

The main issue is how to build the quality function
starting from the expression of the quality of a single community. We have seen 
that there are reliable ways to do it, when one wants to find the best partition in a fixed number of modules. 
Modularity itself, for instance, is a possible prescription.
Instead, the problem of discriminating whether a partition of a network in $n$ modules
is better than a partition in $m$ modules, with $m\neq n$ is more difficult to control, and so far unsolved.
As long as this problem remains open, using the optimization of quality functions
to identify communities will be unjustified.



\begin{thebibliography}{1}  

\bibitem{bara02}
A.-L.~Barab\'{a}si and  R.~Albert,  
``Statistical mechanics of complex networks'',
{\em Rev. Mod. Phys.} {\bf 74}, pp.~47--97, 2002.

\bibitem{mendes03}
S.~N.~Dorogovtsev and J.~F.~F.~Mendes,
{\em Evolution of Networks: from biological nets to the Internet and
WWW}, Oxford University Press, Oxford, UK, 2003.

\bibitem{Newman:2003} M.~E.~J.~Newman, 
``The structure and function of complex networks'',
{\em SIAM Review} {\bf 45}, pp.~167--256, 2003. 

\bibitem{psvbook}
R.~Pastor-Satorras and A.~Vespignani,
{\em Evolution and structure of the Internet: A statistical physics
approach}, Cambridge University Press, Cambridge, UK, 2004.

\bibitem{vitorep} 
S.~Boccaletti, V.~Latora, Y.~Moreno, M.~Chavez and D.-U.~Hwang,
``Complex Networks: Structure and Dynamics'',
{\em Phys. Rep.} {\bf 424}, pp.~175--308, 2006.

\bibitem{Flake:2002} G.~W.~Flake, S.~Lawrence, C.~Lee Giles and F.~M.~Coetzee,
``Self-Organization and Identification of Web Communities'',
{\em IEEE Computer} \textbf{35}(3), pp.~66--71, 2002.

\bibitem{Girvan:2002} M.~Girvan and M.~E.~J.~Newman, 
``Community structure in social and biological networks'',
{\em Proc. Natl. Acad. Sci.} {\bf 99}, pp.~7821--7826, 2002.

\bibitem{Lusseau:2003} D.~Lusseau and M.~E.~J.~Newman, 
``Identifying the role that animals play in their social networks'',
{\em Proc. R. Soc. London B} {\bf 271},pp.~S477--S481, 2004.

\bibitem{Adamic:2005} L.~Adamic and N.~Glance, 
``The Political Blogosphere and the 2004 U.S. Election: Divided They Blog'',
in {\em Proc. $3^{rd}$ Int. Workshop
on Link Discovery}, pp.~36--43, 2005.

\bibitem{foodw1} S.~L.~Pimm, 
``The structure of food webs'',
{\em Theor. Popul. Biol.} \textbf{16}, pp.~144--158, 1979.
 
\bibitem{foodw2} A.~E.~Krause, K.~A.~Frank, D.~M.~Mason, R.~E.~Ulanowicz and W.~W.~Taylor, 
``Compartments exposed in food-web structure'',
{\em Nature} \textbf{426}, pp.~282--285 ,2003.

\bibitem{Newman:2004b} M.~E.~J.~Newman and M.~Girvan, 
``Finding and evaluating community structure in networks'',
{\em Phys. Rev. E} {\bf 69}, 026113, 2004.

\bibitem{brandes} U.~Brandes, D.~Delling, M.~Gaertler, R.~Goerke, M.~Hoefer, Z.~Nikoloski and D.~Wagner, 
``Maximizing modularity is hard'', physics/0608255 in www.arxiv.org.

\bibitem{Newman:2004c} M.~E.~J.~Newman, 
``Fast algorithm for detecting community structure in networks'',
{\em Physical Review E} {\bf 69}, 066133, 2004.

\bibitem{Clauset:2004} A. Clauset, M.~E.~J.~Newman and C. Moore, 
``Finding community structure in very large networks'',
{\em Phys. Rev. E} \textbf{70}, 066111, 2004.

\bibitem{Guimera:2004} R.~Guimer\`a, M.~Sales-Pardo and L.~A.~N.~Amaral, 
``Modularity from fluctuations in random graphs and complex networks'',
{\em Phys. Rev. E} {\bf 70}, 025101(R), 2004.

\bibitem{Guimera:2005} R.~Guimer\`a and L.~A.~N.~Amaral, 
``Functional cartography of complex metabolic networks'',
{\em Nature} {\bf 433}, pp.~895-900, 2005.

\bibitem{Reichardt:2006} J.~Reichardt and S.~Bornholdt, 
``Statistical mechanics of community detection'',
{\em Physical Review E} {\bf 74}, 016110, 2006.

\bibitem{Duch:2005} J.~Duch and A.~Arenas, 
``Community detection in complex networks using extremal optimization'',
{\em Phys. Rev. E} {\bf 72}, 027104, 2005.  

\bibitem{Newman:2006} M.~E.~J.~Newman,
``Modularity and community structure in networks'',
{\em Proc. Natl. Acad. Sci. USA} {\bf 103}, pp.~8577--8582, 2006.

\bibitem{fortunato:2007} S.~Fortunato and M.~Barth\'elemy,
``Resolution limit in community detection'',
{\em Proc. Natl. Acad. Sci. USA} {\bf 104}, pp.~36--41, 2007.

\bibitem{kumpula:2007} J.~M.~Kumpula, J.~Saram\"aki, K.~Kaski and J.~Kert\'esz,
``Limited resolution in complex network community detection with Potts model approach'',
{\em Eur. Phys. J. B} {\bf 56}, pp.~41--45, 2007.

\bibitem{Muff:2005} S.~Muff, F.~Rao and A.~Caflisch, 
`` Local modularity measure for network clusterizations'',
{\em Phys. Rev. E} {\bf 72}, 056107, 2005.

\bibitem{Bornholdt:2006} J.~Reichardt and S.~Bornholdt,
``When are networks truly modular?'',
{\em Physica D} {\bf 224}, pp.~20--26, 2006.

\end{thebibliography}
\end{document}